\documentclass[conference]{IEEEtran}
\IEEEoverridecommandlockouts
\usepackage{amsmath,amssymb,amsfonts}
\usepackage{graphicx}
\usepackage{cite}
\usepackage{booktabs}
\usepackage{url}

\begin{document}

\title{Measurement-Aware Multimodal LLM for No-Reference Low-Dose CT Quality Assessment%
\thanks{This work has been submitted to the IEEE for possible publication. Copyright may be transferred without
notice, after which this version may no longer be accessible.}}

\author{\IEEEauthorblockN{Kerem Basim\textsuperscript{1,*}, Tolga Demiroglu\textsuperscript{1,*},
Mehmet Ozan Unal\textsuperscript{1}, Metin Ertas\textsuperscript{2}, Isa Yildirim\textsuperscript{1}}
\IEEEauthorblockA{\textsuperscript{1}Dept. of Electronics and Communication Eng., Istanbul Technical University,
Istanbul, Turkey\\
\textsuperscript{2}Dept. of Electrical and Electronics Eng., Istanbul University, Istanbul, Turkey\\
\{basim21, unalmehmet, iyildirim\}@itu.edu.tr, tolgadem7@gmail.com, ertas@istanbul.edu.tr\\
\textsuperscript{*}Equal contribution}}

\maketitle

\begin{abstract}
No-reference image quality assessment (IQA) of low-dose CT predicts the score that radiologists would assign to a
slice without a full-dose reference, as needed to compare dose protocols and denoisers. Trained CT IQA networks agree
well with radiologists but output only a score, whereas multimodal large language models (MLLMs) can
explain a score but perform poorly on CT when only prompted. We propose a measurement-aware MLLM, built on Qwen3.8-27B, that rates a
slice from its image together with six image-domain measurements of noise, sharpness, high-frequency content and
streaks, given to the language model as percentiles in text. On the LDCT-IQA 2023 test set, it reaches an Overall score of 2.735, on par with
the retrained challenge winner (2.735) and MedSigLIP-FiLM (2.713), and substantially above its base MLLM used as a
zero-shot agent. Shuffling the measurements at test time reduces the score by 0.048, mostly through noise, but the
same measurements also lead the model to assign higher scores to over-smoothed denoised images.

\end{abstract}


\section{Introduction}
\label{sec:intro}
Lowering the dose of a CT scan reduces radiation risk but increases noise and streak artifacts, and the deep learning
denoisers that remove them can over-smooth fine anatomy. Comparing dose protocols and denoisers therefore requires
image quality assessment (IQA) that agrees with radiologists, and manual scoring by radiologists is impractical at
scale~\cite{lee2025low}. Since no full-dose reference exists in clinical practice, the assessment must be
\emph{no-reference}, that is, a model predicts the mean opinion score (MOS) of radiologists from the slice alone. The
LDCT-IQA 2023 challenge~\cite{lee2025low} established this task as a benchmark of abdominal CT slices scored by
radiologists from 0 (worst) to 4 (best).

On this benchmark, trained networks achieve high correlation with the MOS. The challenge was won by an ensemble of
convolutional and transformer classifiers~\cite{galdran2023ldctiqacode}, later work adapted medical foundation models such as
MedSigLIP-FiLM~\cite{demiroglu2025prompt}, and a recent method reports a higher score~\cite{xiao2026clinreadnet}.
Such scores, however, depend on the evaluation protocol. Rounding predictions to the 0.2 MOS grid changes Kendall's
$\tau_b$, and choosing a model configuration by its test-set score inflates the result. Moreover, these networks
output only a score, whereas multimodal large language models (MLLMs) can also explain a score in words. Prompted
general MLLMs~\cite{celik2025lmm} and a vision--language model paired with ChatGPT~\cite{chen2024iqagpt} assess CT
quality in text, but on LDCT-IQA prompting remains substantially below trained networks (Table~\ref{tab:comparison}).
A medical MLLM fine-tuned for CT quality~\cite{wang2026radiologist} also remains below them. For natural images and videos, MLLMs
are either fine-tuned to output a quality score~\cite{wu2023qalign,wu2025visualqualityr1} or given IQA tools that they
call and reason over in an agent loop~\cite{zhu2025agenticiqa,qin2026tooliqa,wu2026iqat1,xing2025qrouter}.

\begin{figure*}[t]
\centering
\includegraphics[width=0.75\textwidth]{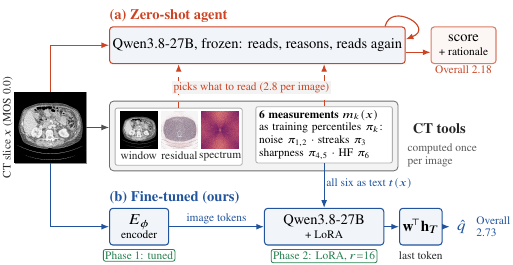}
\caption{Two ways to give Qwen3.8-27B CT quality measurements. (a) Zero-shot agent: it reads precomputed tool
outputs (measurements and visual views) in a loop, 2.8 reads per image on average, and writes a score and a
rationale (Overall 2.18). (b) Our fine-tuned model: six measurement percentiles are written as
text after the image tokens of a CT-tuned vision encoder; LoRA adapters and a linear head on the last token are
trained to predict the MOS (Overall 2.73). Tiles are from a training slice and computed with simple definitions for
illustration.}
\label{fig:method}
\end{figure*}

For CT, the natural counterpart of such tools are the physical measurements with which image quality is
conventionally characterized, such as noise magnitude and edge sharpness~\cite{samei2019tg233,greffier2022iqmetrix},
and an MLLM can read them as text. However, it remains unclear whether an MLLM fine-tuned with these measurements
reaches the accuracy of trained CT IQA networks, and whether it actually uses them. We propose a measurement-aware MLLM
that rates a slice from the image together with six measurements in four groups (noise, sharpness, high-frequency
content and streaks), given as percentiles in text (Fig.~\ref{fig:method}b). We compare this model with its base MLLM, Qwen3.8-27B, used as a zero-shot agent over precomputed measurements and visual views
(Fig.~\ref{fig:method}a). Our contributions are
\textit{(i)}~a measurement-aware MLLM that is on par with the strongest trained CT IQA networks on LDCT-IQA 2023
(Table~\ref{tab:comparison});
\textit{(ii)}~ablation and test-time shuffling experiments showing that the model relies on the measurement values,
mainly on noise, and that the same inputs lead it to assign higher scores to over-smoothed denoised images
(Tables~\ref{tab:ablation} and~\ref{tab:reliance}, Fig.~\ref{fig:interp}); and
\textit{(iii)}~an evaluation under a fixed protocol, with the effect of rounding to the 0.2 MOS grid and
configurations selected on the test set reported separately.

\section{Method}
\label{sec:method}

Given a CT slice $x$, our model (Fig.~\ref{fig:method}b) predicts its MOS $q \in [0,4]$ with a multimodal LLM that
receives the image and six image measurements written as text.

For each slice we compute six scalar measurements $m_k(x)$ that capture the degradations caused by dose reduction and
denoising. They fall into four groups. Noise is measured by the standard deviation of the residual after a
$5{\times}5$ median filter (residual noise) and by a noise score defined below. Sharpness is measured by the 95th
percentile of the Sobel gradient magnitude (structure) and by the mean deviation of this magnitude from its
Gaussian-smoothed version (gradient deviation). Fine texture is the mean absolute high-frequency residual
$|x - G_{3}*x|$ (HF residual), and streaks are the angular non-uniformity of the mid-frequency residual power inside
the body (streak anisotropy). Because the raw values differ in units, we express each as a percentile $\pi_k$. The
percentile is the share of the 1000 training slices with a larger value, so for the noise measurements a higher
$\pi_k$ means less noise than usual. The noise score is the mean of the residual noise, gradient deviation and HF
residual percentiles. No MOS is used in this step. The six two-digit percentiles are inserted into a fixed prompt
(``\ldots{} each number is the percent of reference slices that measure higher \ldots{} Rate the overall image quality
from 0 (worst) to 4 (best).''), which the LLM reads together with the image tokens.

We build the model on Qwen3.8-27B~\cite{qwen38}, whose vision encoder $E_\phi$ converts the image into tokens that a
merger projects into the language model (Fig.~\ref{fig:method}b). Phase~1 adapts this general-purpose encoder to CT.
We fine-tune $E_\phi$ with a small regression head on its pooled tokens and the six percentiles, using the mean
squared error (MSE) on standardized MOS for 10 epochs. Phase~2 keeps $E_\phi$ frozen and trains the language model to
process the image tokens and the measurement text jointly. It updates the merger, LoRA adapters~\cite{hu2022lora} of
rank 16 on every projection of all 64 language-model layers (bf16), and a linear head
$\hat{q} = \mathbf{w}^\top \mathbf{h}_T$ on the hidden state of the last prompt token, again with MSE, for 3 epochs.
The test prediction averages the models of the 5 cross-validation folds.

To test whether the same kind of evidence helps without training, we also run Qwen3.8-27B as a zero-shot agent
(Fig.~\ref{fig:method}a), following AgenticIQA and Tool-IQA~\cite{zhu2025agenticiqa,qin2026tooliqa}. Its tool outputs,
precomputed for each image, are a file of CT measurements with their percentiles among the training
slices and seven visual views (e.g.\ soft-tissue window, noise residual map, Fourier spectrum). The agent opens the
image and these files with file-read calls and writes a score and a short rationale.

\begin{table*}[t]
\caption{Results on the LDCT-IQA 2023 test set (300 slices) and on denoised images ($\bar{\tau}$). All models we ran
are trained on our folds as 5-fold ensembles and evaluated under a protocol fixed in advance (our model's selection:
Sec.~\ref{sec:method}); LMM-IQA and
MedGemma are cited from their papers. Grid: predictions rounded to the 0.2 MOS grid. $\pm$: std over three training
seeds (point values are the original runs; for MedSigLIP-FiLM its best seed). $\Delta$: difference to MedSigLIP-FiLM,
paired bootstrap over the 300 slices. $^\dagger$ Holm-adjusted $p<0.05$ against ours. Bold: best of the rows above
the bottom block. Bottom block: published results and our configurations chosen on the test set, not comparable to the
rows above.}
\label{tab:comparison}
\centering
\setlength{\tabcolsep}{4pt}
\resizebox{\textwidth}{!}{\begin{tabular}{lccccccc}
\toprule
Method & PLCC $\uparrow$ & SROCC $\uparrow$ & KROCC $\uparrow$ & Overall $\uparrow$ & Overall, grid $\uparrow$ & $\Delta$ [95\% CI] & $\bar{\tau}$ $\uparrow$ \\
\midrule
\multicolumn{8}{l}{\textit{Prompted MLLMs}} \\
LMM-IQA, o3~\cite{celik2025lmm} & .8122 & .8180 & .6347 & 2.2649 & - & - & - \\
Qwen3.8-27B agent (ours, Z3) & .7825 & .7859 & .6117 & 2.1802 & 2.1919 & $-$0.533 [$-$0.668, $-$0.415] & - \\
\midrule
\multicolumn{8}{l}{\textit{Fine-tuned MLLMs}} \\
MedGemma-4B QLoRA, own split~\cite{wang2026radiologist} & .791 & .795 & - & - & - & - & - \\
Q-Align~\cite{wu2023qalign} & .9095 & .9055 & .7422 & 2.5572$^\dagger$ & 2.5779 & $-$0.156 [$-$0.210, $-$0.109] & 0.392 \\
\midrule
\multicolumn{8}{l}{\textit{Trained CT IQA}} \\
ResNeXt-50 & .9426 & .9410 & .7989 & 2.6825$^\dagger$ & 2.6968 & $-$0.031 [$-$0.049, $-$0.013] & 0.029 \\
Challenge winner~\cite{lee2025low,galdran2023ldctiqacode} & .9534 & \textbf{.9543} & \textbf{.8275} & \textbf{2.7352} & 2.7363 & +0.022 [$-$0.002, +0.045] & 0.436 \\
MedSigLIP-FiLM~\cite{demiroglu2025prompt} & .9490{\scriptsize$\,\pm$\,.0103} & .9491{\scriptsize$\,\pm$\,.0051} & .8152{\scriptsize$\,\pm$\,.0098} & 2.7132{\scriptsize$\,\pm$\,.0251} & 2.7327 & reference & \textbf{0.480} \\
\textbf{Ours: Qwen3.8-27B + meas.} & \textbf{.9543}{\scriptsize$\,\pm$\,.0015} & .9537{\scriptsize$\,\pm$\,.0015} & .8265{\scriptsize$\,\pm$\,.0031} & 2.7345{\scriptsize$\,\pm$\,.0060} & \textbf{2.7581} & +0.021 [+0.002, +0.041] & 0.140 \\
\midrule
\multicolumn{8}{l}{\textit{Published or chosen on the test set}} \\
Challenge winner, class-grid output~\cite{lee2025low,galdran2023ldctiqacode} & .9491 & .9495 & .8440 & - & 2.7427 & - & - \\
ClinReadNet~\cite{xiao2026clinreadnet} & - & - & - & - & 2.7690 & - & - \\
MedSigLIP-FiLM, test-selected & - & - & - & - & 2.7694 & - & - \\
Ours, test-selected & - & - & - & - & 2.7767 & - & - \\
\bottomrule
\end{tabular}
}
\end{table*}

We evaluate all models with the challenge score~\cite{lee2025low}, Overall $=$ $|$PLCC$|$ $+$ $|$SROCC$|$ $+$
$|$KROCC$|$, the sum of the linear, rank and pairwise-rank correlations with the MOS (KROCC as Kendall's $\tau_b$,
maximum 3). The MOS is the mean of the radiologists' scores and therefore takes values on a 0.2 grid. Predictions can
be rounded to the same grid, $\hat{q} \leftarrow 0.2\,\mathrm{round}(\hat{q}/0.2)$. Rounding creates tied pairs,
which changes KROCC, so we report Overall with and without rounding for every model we ran. To limit selection on the
test set, all comparisons and tests were fixed before the first evaluation on it. Confidence intervals (CIs) come from a paired
bootstrap over the 300 test slices (2000 resamples)~\cite{efron1993bootstrap}, and $p$-values are
Holm-corrected~\cite{holm1979simple} within each table. Since denoised images have no MOS, the denoiser $\bar{\tau}$
measures whether a model ranks the reconstructions of a slice by fidelity. It is the Kendall $\tau$ between predicted
quality and SSIM to the full-dose reference, averaged over slices. One caveat remains. The rule fixed in advance selects by out-of-fold score and picked the Phase~1 model,
whereas the Phase~2 model was promoted after the test evaluation; a rerun with the same seed reproduced its
score.

\section{Experiments and Results}
\label{sec:experiments}

\begin{figure*}[t]
\centering
\includegraphics[width=\textwidth]{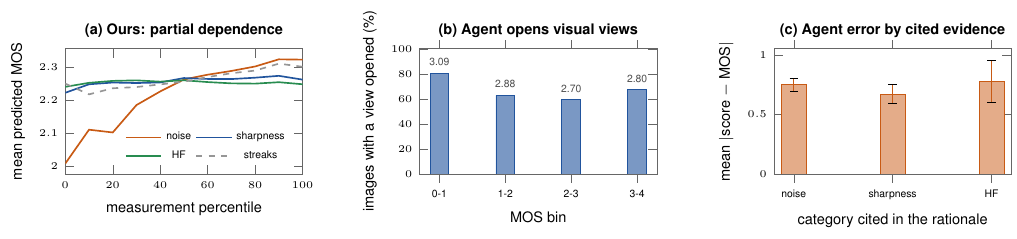}
\caption{What the models rely on. (a) Our model, partial dependence (fold 1): its mean prediction moves mostly with
the noise measurements. (b) The zero-shot agent opens a visual view most often for the worst images (MOS 0--1);
numbers on the bars are reads per image. (c) Agent absolute error by the measurement group its rationale cites
(95\% bootstrap CIs).}
\label{fig:interp}
\end{figure*}

\begin{table}[t]
\caption{Ablation. Each row changes one component, with the same folds, seed and 5-fold ensemble. Meas.: the six
measurement percentiles as text. $\Delta$: against the row above. $^\dagger$ Holm-adjusted $p<0.05$. Bold: our model.}
\label{tab:ablation}
\centering
\setlength{\tabcolsep}{4pt}
\resizebox{\columnwidth}{!}{\begin{tabular}{llccc}
\toprule
& Configuration & Overall $\uparrow$ & $\Delta$ [95\% CI] & $\bar{\tau}$ $\uparrow$ \\
\midrule
\multicolumn{5}{l}{\textit{Zero-shot Qwen3.8-27B (no training)}} \\
Z1 & Image only & 1.370 & - & - \\
Z2 & + meas. as text & 2.157 & +0.787 {\footnotesize [+0.545, +1.028]}$^\dagger$ & - \\
Z3 & + agent loop (files, views) & 2.180 & +0.023 {\footnotesize [$-$0.125, +0.173]} & - \\
Z4 & + 80 examples with MOS & 2.483 & +0.302 {\footnotesize [+0.179, +0.436]}$^\dagger$ & - \\
\midrule
\multicolumn{5}{l}{\textit{Trained, no LLM}} \\
T0 & Meas. only (boosting) & 2.611 & - & 0.056 \\
T1 & Frozen enc. + ridge & 2.569 & - & 0.442 \\
T2 & Tuned enc. & 2.704 & +0.135 {\footnotesize [+0.098, +0.174]}$^\dagger$ & 0.213 \\
T3 & + meas. in head (Phase 1) & 2.719 & +0.015 {\footnotesize [+0.008, +0.022]}$^\dagger$ & 0.150 \\
\midrule
\multicolumn{5}{l}{\textit{Trained, with LLM (LoRA)}} \\
T4 & Frozen enc. + LLM + meas. & 2.711 & - & 0.149 \\
T5 & Tuned enc. + LLM, no meas. & 2.719 & - & 0.250 \\
\textbf{T6} & \textbf{Tuned enc. + LLM + meas.} & \textbf{2.7345} & \textbf{+0.015 {\footnotesize [+0.005, +0.025]}$^\dagger$} & \textbf{0.140} \\
T7 & Ours, meas. shuffled at test & 2.687 & $-$0.048 {\footnotesize [$-$0.070, $-$0.027]}$^\dagger$ & 0.222 \\
\bottomrule
\end{tabular}
}
\end{table}

\begin{table}[t]
\caption{What the models rely on. Ours: change in Overall when one measurement group is shuffled across test images
(5 permutations). Agent: share of images on which it read each group, cited it (frozen keyword rules), and the
citation agreed with the measured percentile. $^\dagger$ Holm-adjusted $p<0.05$. $^{*}$ A human
reader coded noise in 90 to 94\%. $^\ddagger$ Rules not confirmed by a human reader ($\kappa<0.7$).}
\label{tab:reliance}
\centering
\resizebox{\columnwidth}{!}{\begin{tabular}{lcccc}
\toprule
& Ours & \multicolumn{3}{c}{Zero-shot agent (\%)} \\
\cmidrule(lr){2-2}\cmidrule(lr){3-5}
Group & $\Delta$ shuffled [95\% CI] & read & cited & agrees [95\% CI] \\
\midrule
Noise level & $-$0.024 [$-$0.043, $-$0.007]$^\dagger$ & 100 & 99.7$^{*}$ & 76.8 [72.0, 81.9] \\
Sharpness & $-$0.005 [$-$0.012, +0.001] & 100 & 52.3 & 90.3 [85.7, 94.2] \\
HF residual & $-$0.002 [$-$0.007, +0.002] & 100 & 10.3 & 87.1 [74.2, 96.8] \\
Streaks & $-$0.006 [$-$0.014, +0.001] & 100 & 15.0$^\ddagger$ & 80.0 [68.9, 91.1]$^\ddagger$ \\
Visual views & - & 66.0 & - & - \\
All shuffled (T7) & $-$0.048 [$-$0.070, $-$0.027]$^\dagger$ & - & - & - \\
\midrule
Reads per image & - & 2.84 & - & - \\
\bottomrule
\end{tabular}
}
\end{table}

We train and test on LDCT-IQA 2023~\cite{lee2025low}, which provides 1000 training and 300 test abdominal slices
from 7 and 5 patients. We split the training slices (about 100 base slices, each degraded at several levels) into 5
folds grouped by base slice and evaluate on the official test set. Because one patient's
slices can span folds, out-of-fold scores overestimate test performance (Overall gap 0.118, and 0.017 with folds
that also separate likely same-patient slices). For denoised
images, we built a set of 192 soft-tissue slices from 30 DeepLesion patients~\cite{yan2018deeplesion}. Each slice has
its full-dose reference and 12 reconstructions, namely a noisy low-dose version and the outputs of 11 denoisers
(e.g.\ BM3D, RED-CNN, DnCNN), and serves only for $\bar{\tau}$. We train MedSigLIP-FiLM~\cite{demiroglu2025prompt} with its original
code and checkpoints, like our model with three seeds, and retrain the transformer
classifiers of the challenge winner~\cite{lee2025low,galdran2023ldctiqacode} and Q-Align~\cite{wu2023qalign} on our folds with their own recipes, and add a ResNeXt-50 as a standard CNN baseline. The code is available at
\url{https://github.com/itu-biai/agentic-ct-iqa}.

Table~\ref{tab:comparison} shows that our model is on par with the strongest trained CT IQA networks. It reaches an
Overall of 2.735, 0.021 above MedSigLIP-FiLM in the primary comparison (95\% CI [+0.002, +0.041]). This difference is
not significant after Holm correction ($p=0.11$), although our model is ahead at all three seed pairings. The
difference to the retrained challenge winner is close to zero ($-$0.001 [$-$0.018, +0.016]), whereas Q-Align and
ResNeXt-50 are significantly lower. These scores are also sensitive to the evaluation protocol. Rounding to the 0.2
grid adds 0.024 to our model but only 0.001 to the challenge winner, whose 21-class output already lies on the grid.
This rounding effect is larger than the gap between our model and MedSigLIP-FiLM (0.021), so differences between the
top methods on this benchmark are only meaningful when all models are evaluated with the same rounding. Selecting
the configuration on the test set raises MedSigLIP-FiLM further, to 2.769, the score reported for
ClinReadNet~\cite{xiao2026clinreadnet} (Table~\ref{tab:comparison}, bottom). On denoised images, however, the ranking
reverses. Our model reaches $\bar{\tau}=0.140$, against 0.480 for MedSigLIP-FiLM and 0.436 for the winner.

Table~\ref{tab:ablation} isolates the contribution of each component. Without training, the measurement text increases the Overall of Qwen3.8-27B from 1.370 to 2.157, the agent
loop adds no significant gain (2.180), and 80 example slices with their MOS increase it only to 2.483 (rows Z1--Z4).
Among the trained models, tuning the vision encoder gives the largest improvement (T2 vs T1, +0.135), and the
tuned encoder represents noise better than the pretrained one (linear-probe $R^2$ 0.91 vs 0.83). Once the LLM receives the measurements, however, the advantage of the tuned over a frozen encoder is no longer
significant (T6 vs T4). The measurement text itself adds 0.015 (T6 vs T5), and shuffling it across test images more
than offsets this gain (T7, $-$0.048). The same text also explains the poor ranking of denoised images. It lowers
$\bar{\tau}$ from 0.250 to 0.140, and shuffling it restores $\bar{\tau}$ to 0.222.

Among the measurements, our model relies mainly on noise. Shuffling one group at a time significantly reduces Overall
only for noise ($-$0.024, Table~\ref{tab:reliance}), and the mean prediction changes by 0.32 MOS across the noise
percentiles but by at most 0.09 for the other groups (Fig.~\ref{fig:interp}a). The zero-shot agent also relies on
noise, but less consistently. It reads the measurement file for every image and opens a visual view for 66\%, most
often for the worst images (Fig.~\ref{fig:interp}b). Its rationales cite noise almost always, yet the cited direction
agrees with the measured percentile for only 77\% of images. These citations were counted with keyword rules fixed in
advance, which a human reader confirmed for sharpness and HF ($\kappa=0.88$) and for the direction of noise citations
(96\%), but not for streaks. The agent's
error is similar across the cited groups (overlapping 95\% CIs, Fig.~\ref{fig:interp}c), and removing the noise tools
changes 41\% of its scores by more than 0.25 MOS without changing its Overall. The rationales therefore describe the
measurements but do not reliably explain the scores.

\section{Conclusion}
\label{sec:conclusion}
A measurement-aware MLLM that reads six image measurements as text together with the CT image is on par with the
strongest trained networks on LDCT-IQA 2023, whereas its base MLLM as a zero-shot agent scores substantially lower.
The model relies mainly on the noise measurements, which also lead it to overrate over-smoothed
denoised images and limit its use for ranking denoisers. The study is limited to 7 training patients, one
benchmark and one LLM, and the denoised set has no MOS. Future work will add training images with denoiser blur and
evaluate smaller MLLMs.

\bibliographystyle{IEEEtran}
\bibliography{refs}

\end{document}